\documentclass[prd,showpacs,preprintnumbers,twocolumn,amsmath,nofootinbib,amssymb]{revtex4}
\usepackage{graphicx,color,dcolumn,booktabs,bm}
\usepackage{longtable,lscape}
\usepackage{txfonts}
\usepackage{overpic}
\usepackage{amssymb}
\usepackage{epstopdf}
\usepackage{indentfirst}
\usepackage{feynmf}   
\usepackage{slashed}  
\usepackage{cases}
\usepackage{color}
\usepackage{float}
\usepackage{multirow}
\usepackage{graphicx,color,dcolumn,booktabs,bm}
\usepackage{epsfig,dsfont,amssymb,amsmath,amsfonts,amsbsy,mathrsfs}

\graphicspath{{Figures/}} %
\usepackage[colorlinks, citecolor=blue,anchorcolor=red,menucolor=red, linkcolor=red,filecolor=red,runcolor=red,urlcolor=blue,frenchlinks=red]{hyperref}

\makeatletter
\@addtoreset{equation}{section}
\makeatother

\allowdisplaybreaks
\begin{document}
\title{Excited states of $\phi$  meson}
\author{Cheng-Qun Pang$^{1}$}
\email{xuehua45@163.com}
\affiliation{$^1$College of Physics and Electronic Information Engineering, Qinghai Normal University, Xining 810000, China\\}
\begin{abstract}
In this paper, the excited states of the $\phi$  meson, especially containing the newly observed $X(2000)$ with $I(J^P)=0(1^-)$ by the BESIII Collaboration, is studied. In addition, $Y(2175)$ as a  $\phi$ {\color{black}{meson excited state}} is  investigated.
The mass spectrum and strong decay behaviors of $\phi$ {\color{black}{meson excited states}} are analyzed, which indicates that $X(2000)$ and $Y(2175)$ are the candidates of $\phi(3S)$ and $\phi(2D)$ states with $I(J^P)=0(1^-)$, respectively. In addition,  {\color{black}{$\phi(1D)$ and $\phi(4S)$ are}} predicted to have the mass of 1.87 GeV and 2.5 GeV and width of 550 MeV and 230 MeV, respectively.
\end{abstract}
\maketitle

\section{Introduction}\label{sec1}
{\color{black}{In light meson spectroscopy, there exist two systems: $q\bar q$ ($q$ defines the $u,d$ or $s$ quark) mesons and exotic states (glueballs, hybrids, and multiquark states).
Exotic states have exotic quantum
numbers (such as $\pi_1(1400)$) or the same quantum numbers as the conventional meson system. In the latter case, i.e.,
if exotic states have the same quantum numbers as conventional meson system, it is  difficult but intriguing to identify the  exotica from  light meson spectroscopy \cite{Li:2005bz,Ding:2005ew,Wang:2010vz,Li:2006ru,Chao:2006fq,Bicudo:2006sd,Klempt:2007cp,Wang:2006ri,Ding:2006ya,MartinezTorres:2008gy,Wang:2017iai,Huang:2005bc,Yu:2011ta,Liu:2010tr,Li:2008mza,Wang:2012wa,Ding:2007pc}.
Thus, some states may be the conventional $q\bar{q}$ mesons or exotic states, which are related to $X(1835)$ \cite{Ablikim:2005um,Ablikim:2010au}, $X(1860)$ \cite{Bai:2003sw}, $X(1812)$, $Y(2175)$ \cite{Shen:2009zze,Aubert:2006bu,Ablikim:2018xuz} and so on}}.
\par
Very recently, the {\color{black}{BESIII Collaboration observed}} a structure ($X(2000)$) appearing in the $\phi \eta^{\prime}$ invariant mass spectrum of the $J/\psi \to \phi\eta\eta^{\prime}$ process \cite{Ablikim:2018xuz}. With assumption that  the spin-parity quantum number $J^P=1^-$, its measured resonance parameters are M=$2002.1\pm27.5\pm15.0\,{\text{MeV}}$ and
$\Gamma=129\pm17\pm7\,{\text{MeV}}$ with significance of 5.3$\sigma$.
Assuming the spin-parity quantum number   $J^P=1^+$, the resonance {\color{black}{parameters are}}   M=$2062.8\pm13.1\pm4.2\,{\text{MeV}}$ $\Gamma=177\pm36\pm20\,{\text{MeV}}$ with significance of 4.9$\sigma$.

Naturally, one can note that the resonance parameters with the first assumption
are close to the observed $Y(2175)$ \cite{Shen:2009zze,Aubert:2006bu,Ablikim:2018xuz}. $Y(2175)$ has been studied  in various theoretical explanations \cite{Li:2005bz,Ding:2005ew,Wang:2010vz,Li:2006ru,
Chao:2006fq,Bicudo:2006sd,Klempt:2007cp,Wang:2006ri,Ding:2006ya,MartinezTorres:2008gy,Wang:2017iai,Huang:2005bc,Yu:2011ta,Liu:2010tr,Li:2008mza,Wang:2012wa,Ding:2007pc}. $X(2000)$ {\color{black}{has also been}} studied by recent {\color{black}{work}} \cite{Wang:2019qyy,Cui:2019roq,Wang:2019nln}.
Reference \cite{Wang:2012wa} studied  $Y(2175)$ as a $\phi(2^3D_1)$ state. Reference \cite{Wang:2019qyy} treated $X(2000)$ as a $h_1(3^1P_1)$ state with  $s\bar s$ component under $J^P=1^+$ assignment,  {\color{black}{Cui ${\it et~al}$. \cite{Cui:2019roq} argued that the $X(2000)$  is the partner of the tetraquark state $Y(2175)$ with $J^{P} = 1^{+}$, and Ref. \cite{Wang:2019nln}  assigned $X(2000)$  to be a new $ss\bar{s}\bar{s}$ tetraquark state with the same $J^{P}$.
As another possibility, i.e.,  $X(2000)$ has the resonance parameters M=$2002.1\pm27.5\pm15.0\,{\text{MeV}}$ and
$\Gamma=129\pm17\pm7\,{\text{MeV}}$ with $J^{P} = 1^{-}$, has not been theoretically studied.
In addition,  a hybrid with the same quantum numbers and a similar mass and width are predicted by the flux-tube model \cite{Isgur:1985vy,Barnes:1995hc,Page:1998gz,Close:1994hc}.
Identifying whether $X(2000)$ is $s\bar{s}$ or $s\bar{s}$ hybrid is a difficult, interesting, and urgent research issue.
In $J^{PC}=1^{--}$ assignment, $X(2000)$ is the candidate of an excited state of $\phi$ meson in the conventional  $s\bar{s}$ meson framework. In fact, }}
Refs. \cite{Barnes:2002mu,Afonin:2014nya} predicted a $\phi(3S)$ state with the mass of 2050 MeV and 1900-1960 MeV, Refs. \cite{Barnes:2002mu} also predicted that the width of $\phi(3S)$ will be 380 MeV. If  this $X$ state is considered as the conventional mesons under the $J^P=1^-$ assignment, what is the relation between $X(2000)$ and $Y(2175)$? Is $X(2000)$ a $\phi(3S)$ state? These questions should be clarified.
In addition, the angular excited state of $\phi(1S)$, the mass and the width of $\phi(1D)$  {\color{black}{are}} unclear. {\color{black}{A systemic study of excited states of $\phi$ meson represents}}  an intriguing and important research topic.
\par
This paper is aimed to give a {\color{black}{systemic study of excited states of $\phi$ meson}}. By using modified Godfrey-Isgur (GI) model and quark pair creation model, the mass spectrum and strong decay  {\color{black}{behavior of excited states of $\phi$  meson }} are analyzed, which indicates  that $X(2000)$  is a candidate  {\color{black}{of}} the $\phi(3S)$ meson with $I(J^P)=0(1^-)$ and $Y(2175)$ is a candidate of the $\phi(2D)$ state. At the same time, the mass and the width of $\phi(1D)$, $\phi(3D)$, and $\phi(4S)$ are predicted.
\par

In this work, the  spectra of the  $\phi$  meson family {\color{black}{are studied}} using the modified  Godfrey-Isgur (MGI) model \cite{Song:2015nia,Song:2015fha,Pang:2017dlw,Pang:2018gcn},
which contains the screening effect. At higher excited states of $\phi$ meson, the screening effect should be considered for the larger average distance between the quark pair.
  The former studies \cite{Godfrey:1985xj,Barnes:2005pb,Sun:2014wea,Song:2015nia,Song:2015fha,Godfrey:2015dia,Godfrey:2016nwn,Pang:2017dlw} show that the GI model works well for describing hadron spectroscopy. Then, for further studying the properties of   $\phi$ mesons,  their Okubo-Zweig-Iizuka (OZI)-allowed two-body strong decays {\color{black}{are studied}},  taking input with the spatial wave functions obtained from the mass spectrum {\color{black}{by numerical calculation}}.
Their partial and total decay widths are calculated by using a quark pair creation (QPC) model that was proposed by Micu \cite{Micu:1968mk} and extensively applied to studies of strong decay of other hadrons \cite{LeYaouanc:1972ae,
vanBeveren:1979bd,vanBeveren:1982qb,LeYaouanc:1988fx,roberts,Capstick:1993kb,Blundell:1995ev,Ackleh:1996yt,Capstick:1996ib,Bonnaz:2001aj,
Close:2005se,Zhang:2006yj,Lu:2006ry,Sun:2009tg,Liu:2009fe,Sun:2010pg,Rijken:2010zza,Yu:2011ta,Zhou:2011sp,Ye:2012gu,Wang:2012wa,
Sun:2013qca,He:2013ttg,Sun:2014wea,Pang:2014laa,Wang:2014sea,Chen:2015iqa,Pang:2017dlw,Pang:2018gcn,Pan:2016bac}. This paper also gives a comparison of  $X(2000)^\prime$s two-body decay  information between that of $\phi(3S)$  and $s\bar{s}$ hybrid  \cite{Page:1998gz}.
 The effort will be helpful to uncover the structure of $X(2000)$ and $Y(2175)$ and establish $\phi$ meson families.
\par
This paper is organized as follows. In Sec. \ref{sec2}, the models employed in this work are briefly reviewed.
The mass spectrum and decay behavior phenomenological analysis of $\phi$ mesons will be performed in {\color{black}{Sec.\ref{sec3}}. The paper ends with a  conclusion in Sec. \ref{sec4}.
\section{Models employed in the work}\label{sec2}

In this work, the modified GI quark model and quark pair creation (QPC) model are utilized to calculate the mass spectrum and the two-body strong decays of the  meson family, respectively.
In the following, these models will be illustrated briefly.

\subsection{The modified GI model}

In 1985, Godfrey and {Isgur} proposed the GI model for describing relativistic meson spectra with great success, specifically for low-lying mesons\cite{Godfrey:1985xj}.
Regarding the excited states, the screening potential should be taken into account for its coupled-channel effect \cite{Song:2015nia,Song:2015fha,Pang:2017dlw,Pang:2018gcn}.

The internal interaction of mesons is depicted by the Hamiltonian of the potential model and can be written as
\begin{equation}\label{Hamtn}
  \tilde{H}=\sqrt{m_1^2+\mathbf{p}^2}+\sqrt{m_2^2+\mathbf{p}^2}+\tilde{V}_{\mathrm{eff}}(\mathbf{p,r}),
\end{equation}
where $m_1$ and $m_2$ denote the mass of quark and antiquark, respectively, {\color{black}{the relation between $\tilde{V}_{\mathrm{eff}}(\mathbf{p,r})$ and $V_{\mathrm{eff}}(\mathbf{p,r})$ will be illustrated later}}, and the
 effective potential has a familiar format in the nonrelativistic limit \cite{Godfrey:1985xj,Lucha:1991vn}
\begin{eqnarray}
V_{\mathrm{eff}}(r)=H^{\mathrm{conf}}+H^{\mathrm{hyp}}+H^{\mathrm{so}}\label{1},
\end{eqnarray}
with
\begin{align}
 H^{\mathrm{conf}}&=\Big[-\frac{3}{4}(\frac{b(1-e^{-\mu r})}{\mu}+c)+\frac{\alpha_s(r)}{r}\Big](\bm{F}_1\cdot\bm{F}_2)\nonumber\\ &=S(r)+G(r),\label{3}\\
H^{\mathrm{hyp}}&=-\frac{\alpha_s(r)}{m_{1}m_{2}}\Bigg[\frac{8\pi}{3}\bm{S}_1\cdot\bm{S}_2\delta^3 (\bm r) +\frac{1}{r^3}\Big(\frac{3\bm{S}_1\cdot\bm r \bm{S}_2\cdot\bm r}{r^2} \nonumber  \\ \label{3.1}
&\quad -\bm{S}_1\cdot\bm{S}_2\Big)\Bigg] (\bm{F}_1\cdot\bm{F}_2), \\
H^{\mathrm{so}}=&H^{\mathrm{so(cm)}}+H^{\mathrm{so(tp)}},  \label{3.2}
\end{align}
where $\bm{S}_1/\bm{S}_2$ indicates the spin of quark/antiquark and $\bm{L}$ is the orbital momentum. {\color{black}{$\bm{F}$ are related to the Gell-Mann matrices in color space}}. For a meson, $\langle\bm{F}_1\cdot\bm{F}_2 \rangle=-4/3$, the running coupling constant $\alpha_s(r)$ has the following form:
 \begin{align}
  \alpha_s(r)=\sum_{k}\frac{2\alpha_k}{\sqrt{\pi}} \int_{0}^{\gamma_k r}e^{-x^2}dx,
 \end{align}
 where $k$ is from 1 to 3 and the corresponding $\alpha_k$ and $\gamma_k$ are constant, $\alpha_{1,2,3}=0.25,0.15,0.2$  and $\gamma_{1.2.3}=\frac{1}{2},\frac{\sqrt{10}}{2},\frac{\sqrt{1000}}{2}$ \cite{Godfrey:1985xj}. $H^{\mathrm{conf}}$ consists of two pieces, the spin-independent linear confinement piece $S(r)$ and Coulomb-like potential $G(r)$. $H^{\mathrm{hyp}}$ is the color-hyperfine interaction and also includes two parts, tensor and contact terms; $H^{\mathrm{SO}}$ denotes the spin-orbit interaction
 with the color magnetic term due to one-gluon exchange and the Thomas precession term, which can be written as
\begin{eqnarray}
H^{\mathrm{so(cm)}}=\frac{-\alpha_s(r)}{r^3}\left(\frac{1}{m_{1}}+\frac{1}{m_{2}}\right)\left(\frac{\bm{S}_1}{m_{1}}+\frac{\bm{S}_2}{m_{2}}\right)
\cdot
\bm{L}(\bm{F}_1\cdot\bm{F}_2),
\end{eqnarray}
\begin{eqnarray}
H^{\mathrm{so(tp)}}=-\frac{1}{2r}\frac{\partial H^{\mathrm{conf}}}{\partial
r}\Bigg(\frac{\bm{S}_1}{m^2_{1}}+\frac{\bm{S}_2}{m^2_{2}}\Bigg)\cdot \bm{L}.
\end{eqnarray}

\par
     In the light meson system, relativistic effects in effective potential must be considered; the GI model introduces these relativistic effects in two ways.
 \par
     First, the GI model introduces a smearing  function for a $q\bar{q}$ meson, which includes nonlocal interactions and new $\mathbf{r}$ dependence.
\begin{equation}
\rho \left(\mathbf{r}-\mathbf{r'}\right)=\frac{\sigma^3}{\pi ^{3/2}}e^{-\sigma^2\left(\mathbf{r}-\mathbf{r'}\right)^2},
\end{equation}
then, $S(r)$ and $G(r)$ become smeared potentials $\tilde{S}(r)$ and $\tilde{G}(r)$, respectively, by the following procedure:
\begin{equation}\label{smear}
\tilde{f}(r)=\int d^3r'\rho(\mathbf{r}-\mathbf{r'})f(r'),
\end{equation}
with
\begin{eqnarray}
   \sigma_{12}^2=\sigma_0^2\Bigg[\frac{1}{2}+\frac{1}{2}\left(\frac{4m_1m_2}{(m_1+m_2)^2}\right)^4\Bigg]+
  s^2\left(\frac{2m_1m_2}{m_1+m_2}\right)^2,
\end{eqnarray}
where  the values of $\sigma_0$ and $s$ are defined in Table \ref{SGIfit1} \cite{Pang:2018gcn}.
\renewcommand{\arraystretch}{1.2}
\begin{table}[htbp]
\caption{Parameters and their values in this work, which are determined by fitting  the meson experimental data listed in PDG. \label{SGIfit1}}
\begin{center}
\begin{tabular}{cccc}
\toprule[1pt]\toprule[1pt]
Parameter &  Value \cite{Pang:2018gcn}&Parameter &  Value \cite{Pang:2018gcn} \\
 \midrule[1pt]
$m_u$ (GeV) &0.163&{$\sigma_0$ (GeV)}&{1.799}\\
$m_d$ (GeV) &0.163&{$s$ (GeV)}&{1.497}\\
$m_s$ (GeV) &0.387&$\mu$ (GeV)&0.0635 \\
$b$ (GeV$^2$) &0.221&$c$ (GeV)&-0.240\\
$\epsilon_c$&-0.138&$\epsilon_{sov}$&0.157\\
$\epsilon_{sos}$&0.9726& $\epsilon_t$&0.893\\
\bottomrule[1pt]\bottomrule[1pt]
\end{tabular}
\end{center}
\end{table}

Second, to  make up for the loss  of relativistic effects in the nonrelativistic limit, a general potential relying on the {\color{black}{the center-of-mass}} of interacting quarks and  momentum are applied as
\begin{equation}
\tilde{G}(r)\to \left(1+\frac{p^2}{E_1E_2}\right)^{1/2}\tilde{G}(r)\left(1 +\frac{p^2}{E_1E_2}\right)^{1/2},
\end{equation}
and
\begin{equation}
  \frac{\tilde{V}_i(r)}{m_1m_2}\to \left(\frac{m_1m_2}{E_1E_2}\right)^{1/2+\epsilon_i} \frac{\tilde{V}_i(r)}{m_1 m_2} \left( \frac{m_1 m_2}{E_1 E_2}\right)^{1/2+\epsilon_i},
\end{equation}
where $\tilde{V}_i(r)$  delegates the contact, tensor, vector spin-orbit and scalar spin-orbit terms, and $\epsilon_i$ represents the relevant modification parameters as shown in Table \ref{SGIfit1}. {\color{black}{After the above revision in two points, $\tilde{V}_{\mathrm{eff}}(\mathbf{p,r})$ is replaced by $V_{\mathrm{eff}}(\mathbf{p,r})$}}.

Diagonalizing and solving the Hamiltonian in Eq.(\ref{Hamtn}) by exploiting a simple harmonic oscillator (SHO) basis, the mass spectrum and wave functions will be obtained.
In configuration and momentum space, SHO wave functions have explicit {forms}:
\begin{align}
\Psi_{nLM_L}(\mathbf{r})=R_{nL}(r, \beta)Y_{LM_L}(\Omega_r),\nonumber\\
\Psi_{nLM_L}(\mathbf{p})=R_{nL}(p, \beta)Y_{LM_L}(\Omega_p),
\end{align}
with
\begin{eqnarray}
&R_{nL}(r,\beta)=\beta^{3/2}\sqrt{\frac{2n!}{\Gamma(n+L+3/2)}}(\beta r)^{L}
e^{\frac{-r^2 \beta^2}{2}} \nonumber \\
 &\times L_{n}^{L+1/2}(\beta^2r^2),\\
 &R_{nL}(p,\beta)=\frac{(-1)^n(-i)^L}{ \beta ^{3/2}}e^{-\frac{p^2}{2 \beta ^2}}\sqrt{\frac{2n!}{\Gamma(n+L+3/2)}}{(\frac{p}{\beta})}^{L} \nonumber \\
 &\times L_{n}^{L+1/2}(\frac{p^2}{ \beta ^2}),
\end{eqnarray}
where $Y_{LM_L}(\mathrm{\Omega})$ is spherical harmonic function, $L_{n-1}^{L+1/2}(x)$ is the associated Laguerre polynomial, and $\beta=0.4~\mathrm{GeV}$ for the calculation.

{After diagonalization of the Hamiltonian matrix, the mass and wave function of the meson that is available to undergo the strong decay process can be obtained.}

\subsection{QPC {model}}

{The QPC model is used to obtain the Okubo-Zweig-Iizuka (OZI) allowed hadronic strong decays.
This model was first proposed by Micu \cite{Micu:1968mk}} and was further developed by Orsay group\cite{LeYaouanc:1972ae,LeYaouanc:1973xz,LeYaouanc:1974mr,LeYaouanc:1977gm,LeYaouanc:1977ux}.
The QPC model was widely applied to the OZI-allowed two-body {\color{black}{strong decays}} of hadrons in Refs. \cite{vanBeveren:1979bd,vanBeveren:1982qb,Capstick:1993kb,Page:1995rh,Titov:1995si,Ackleh:1996yt,Blundell:1996as,
Bonnaz:2001aj,Zhou:2004mw,Lu:2006ry,Zhang:2006yj,Luo:2009wu,Sun:2009tg,Liu:2009fe,Sun:2010pg,Rijken:2010zza,Ye:2012gu,
Wang:2012wa,He:2013ttg,Sun:2013qca,Pang:2014laa,Wang:2014sea,Chen:2015iqa,Pang:2017dlw,Pang:2018gcn}.

For the process $A\to B+C$,
\begin{eqnarray}
\langle BC|\mathcal{T}|A \rangle = \delta ^3(\mathbf{P}_B+\mathbf{P}_C)\mathcal{M}^{{M}_{J_{A}}M_{J_{B}}M_{J_{C}}},
\end{eqnarray}
where $\mathbf{P}_{B(C)}$ is a three-momentum of a meson $B(C)$ in the rest frame of a meson $A$. $M_{J_{i}}\, (i=A,B,C)$ {\color{black}{denotes the
magnetic quantum number}}. The transition operator $\mathcal{T}$ describes a quark-antiquark pair creation from vacuum with
$J^{PC}=0^{++}$, i.e., $\mathcal{T}$ can be written as
\begin{eqnarray}
\mathcal{T}& = &-3\gamma \sum_{m}\langle 1m;1~-m|00\rangle\int d \mathbf{p}_3d\mathbf{p}_4\delta ^3 (\mathbf{p}_3+\mathbf{p}_4) \nonumber \\
 && \times \mathcal{Y}_{1m}\left(\frac{\textbf{p}_3-\mathbf{p}_4}{2}\right)\chi _{1,-m}^{34}\phi _{0}^{34}
\left(\omega_{0}^{34}\right)_{ij}b_{3i}^{\dag}(\mathbf{p}_3)d_{4j}^{\dag}(\mathbf{p}_4).
\end{eqnarray}
 where the quark and antiquark are denoted by indices $3$ and $4$, respectively, and $\gamma$ depicts the strength of the creation of $q\bar{q}$ from vacuum. In this work,
 $\gamma=6.57$, which is obtained by fitting the decay width of $\phi(1680)(2S)$ state as shown in Table \ref{decayfit} and is independent of the decay channels$\prime$ branch ratios. $\mathcal{Y}_{\ell m}(\mathbf{p})={|\mathbf{p}|^{\ell}}Y_{\ell
m}(\mathbf{p})$ are the solid harmonics. $\chi$, $\phi$, and $\omega$ denote the spin, flavor, and color wave functions, respectively, which can be separately treated.
Subindices $i$ and $j$ denote the color of a $q\bar{q}$ pair.
The decay width reads
\begin{eqnarray}
\Gamma&=&\frac{\pi}{4} \frac{|\mathbf{P}|}{m_A^2}\sum_{J,L}|\mathcal{M}^{JL}(\mathbf{P})|^2,
\end{eqnarray}
where $m_{A}$ is the mass of an initial state $A$ and
the two decay amplitudes can related to the Jacob-Wick formula as \cite{Jacob:1959at}
\begin{eqnarray}
\mathcal{M}^{JL}(\mathbf{P})&=&\frac{\sqrt{4\pi(2L+1)}}{2J_A+1}\sum_{M_{J_B}M_{J_C}}\langle L0;JM_{J_A}|J_AM_{J_A}\rangle \nonumber \\
&&\times \langle J_BM_{J_B};J_CM_{J_C}|{J_A}M_{J_A}\rangle \mathcal{M}^{M_{J_{A}}M_{J_B}M_{J_C}}.
\end{eqnarray}

In the calculation, the spatial wave functions of the discussed mesons can be numerically obtained by the MGI model.

\begin{table}[htb]
\centering%
\caption{The decay widths of  the $\phi(1680)(2S)$ state. \label{decayfit}}
\begin{tabular}{lcccccccc}
\hline
\hline
   Decay channel&       Expe. (MeV) &      This work\\
\hline
$Total{}$&   150 &150   \\
$\phi(1680)\rightarrow KK^*$&                --&117   \\
$\phi(1680)\rightarrow \eta\phi$&                --&16.7   \\
$\phi(1680)\rightarrow KK$&                --&15.5  \\
\hline
$\gamma =6.57$ &    &\\
\hline
\hline
\end{tabular}
\end{table}
\section{Numerical results and phenomenological analysis}\label{sec3}
\subsection{Mass spectrum analysis}
Applying the MGI model and the parameters in Table \ref{SGIfit1}, the mass spectrum of the $\phi$ family can be obtained, as shown in Table \ref{mass}.
In addition, the mass spectrum of mesons with $J^{P}=1^-$ was calculated {\color{black}{by the GI model,}} { Ref. \cite{Ebert:2009ub} also gave a  spectrum for the $\phi$ meson. The mass spectrum of these $\phi$ states  can  be obtained  by the MGI {\color{black}{model}} which is listed in {Table} \ref{mass}. The numerical results are compared with the GI model \cite{Godfrey:1985xj}, Ref. \cite{Ebert:2009ub}} and experiments in {Table} \ref{mass}.
\begin{table}[htbp]
\caption{The mass spectrum of $\phi$ mesons. "Expe." represents experimental value.
The unit of the mass is GeV. \label{mass}}
\begin{center}
\[\begin{array}{cccccccc}
\hline
\hline
 \text{State} & \text{This work} & \text{GI \cite{Godfrey:1985xj}} &\text{Ebert \cite{Ebert:2009ub}} &\text{Expe.}\\
 \hline
 \phi(1S) & {1.030} & 1.016& 1.038&1.019   \\
 \phi(2S) & {1.687} & 1.687& 1.698&1.680   \\
 \phi(3S) & {2.149} &2.200& 2.119&--   \\
 \phi(4S) & {2.498} &2.622& 2.472&--   \\
 \phi(1D) & {1.869} & 1.876& 1.845&--   \\
 \phi(2D) & {2.276} &2.337& 2.258&--   \\
 \phi(3D) & {2.593} &2.725& 2.607&--   \\
\hline
\hline
\end{array}\]
\end{center}
\end{table}
\par
\subsubsection{The spectrum of $\phi$ meson excitations}
 The spectrum of  $\phi$ meson excitations is calculated, and the values are listed in {Table} \ref{mass}.
 The third radial excited state of $\phi(1S)$  has a mass of 2.5 GeV, which is {\color{black}{smaller}} than the result of GI model and close to that reported in Ref. \cite{Ebert:2009ub}. For the ground  state of a D-wave $\phi$ meson ($\phi(1D)$), its first and second radial excited states  ($\phi(2D)$ and  $\phi(3D)$) have the mass {of 1.869 GeV, 2.276 GeV} and  2.6 GeV, respectively, which are also smaller than those reported in Ref. \cite{Godfrey:1985xj}.

\subsubsection{$Y(2175)$ and $X(2000)$}
According to  {Table} \ref{mass}, one can note that
$Y(2175)$  tends to be the candidate of $\phi(3S)$ rather than $\phi(2D)$ state;
Ref. \cite{Ebert:2009ub} and the MGI model mass spectrum show that $Y(2175)$  could be the  $\phi(3S)$ or $\phi(2D)$  state because the mass of $Y(2175)$  is between their masses.
The position of $Y(2175)$ in the $\phi$ family needs further discussion based on the decay behavior, which will be given in the next section.

As shown in {Table} \ref{mass}, the mass spectrum of Ref. \cite{Ebert:2009ub}, the GI model and MGI model all indicate that the newly observed state
     $X(2000)$ \cite{Ablikim:2018xuz} may be $\phi(1D)$  or $\phi(3S)$ state. In fact, Ref. \cite{Barnes:2002mu} estimated the mass of $\phi(3S)$ to be 2050 MeV, which is smaller than the mass obtained with the GI model, Ref.  \cite{Ebert:2009ub} and MGI model.
Further discussion based on the decay behaviors on the assignment of  $X(2000)$ will be given below.

The above discussions are only from the point of view of the mass spectra.
In the next section, their strong decays will be studied.

\subsection{Decay behavior analysis}

Applying the QPC model, one can obtain  the OZI-allowed two-body strong decay  of vector light family, which is shown in Tables \ref{decay1} and \ref{decay2}.
\begin{table}[htb]
\centering%
\caption{The partial decay widths of the $\phi(3S)$, $s\bar{s}$ hybrid and $\phi(4S)$, the unit of  widths is MeV.   \label{decay1}}
\[\begin{array}{ccc|c|cc}
\hline
\hline
\multicolumn{3}{c|}{\phi(\text{3S})}&s\bar{s}~hybrid &\multicolumn{2}{c}{{\phi \text{(4S)}}}\\%
\hline
mass&2188&2002&2000 $\cite{Page:1998gz}$&\multicolumn{2}{c}{2498}\\%
\midrule[1pt]
Channel &Value &Value&Value $\cite{Page:1998gz}$&Channel &Value\\
Total&{225}&\color{black}{139}&120&Total&{232}\\
KK^*(1410)&{48.4 } &{45.2 }&9&KK^* &{34.3} \\
KK^*  &{60}   & {26.9}& 16 &K^*K_1^\prime&{29.8} \\
KK_1^\prime &{4.36} & {21.4}      &64& KK_2^*(1430) &{26}\\
KK_1   &{31.4}  & {25.7}   &26& KK^*(1410) & {25.3} \\
KK &{11.9}& {10.8} &-& KK_1&{30.0} \\
KK_2^*(1430)& {39.8}& {4.48 } &2 &K^*K_0^*(1425) &{14.0} \\
 K^*K^*&{ 22.7}    & {4.40}  &-&K^*K_1  &{15.1} \\
  \eta\phi & {6.66}& {0.363}  &3&K^*K^* &{12.2 }\\
   \eta^\prime\phi& 0.0862&0.0729&0.02&KK&{9.69}\\
    & &&   &  K^*K^*(1410)  &{9.62} \\
 & { }& &  &  KK_3^*(1780) &{9.72} \\
  & { }& &  &  K^*K_2^*(1430) &{4.75} \\
    & { }& &  &  KK^*(1680) &{4.75} \\
            & { }& &  &  KK_1^\prime &{2.33} \\
       & { }& &  &  f_1(1420)\phi &{2.01} \\
        & { }& &  &  \eta\phi &{1.69} \\


 \hline
 \hline
\end{array}\]
\end{table}
\subsubsection{{\color{black}{The  radial excited states of S-wave $\phi$ meson}}}

In this section, the  radial excited states of S-wave $\phi$ meson   will be discussed.
\par
$\phi(1680)$  has been established as a $\phi(2S)$ state\cite{Barnes:2002mu}. As presented in Table \ref{decayfit}, the branch ratio  $\Gamma_{KK}/\Gamma_{KK^*}$ is approximately 0.13, which is closer to the experimental value $0.07\pm 0.01$ \cite{Mane:1982si} than the theoretical result of Ref. \cite{Barnes:2002mu}. The ratio $\Gamma_{\eta\phi}/\Gamma_{KK^*}$ is predicted to  be 0.14, which is close to the value (0.18) of Ref. \cite{Barnes:2002mu}.

\par
The  decay widths of  $\phi(3S)$ state with the mass of 2188($Y(2175)$) and 2002($X(2000)$) are compared  in Table \ref{decay1}.
Reference \cite{Barnes:2002mu} also estimated the mass and the width of $\phi(3S)$ to be 2050 MeV and 380 MeV, respectively.
 If $Y(2175)$ is the second excited state of $\phi(1S)$, its total width is 225 MeV, which {\color{black}{does not}}  agree with the experimental value \cite{Aubert:2006bu}.
 According to the mass spectrum analysis section and Ref. \cite{Barnes:2002mu}, the mass and the width of $Y(2175)$ will be larger than the theoretical result when it is treated as $\phi(3S)$.
\par

 According to Table \ref{decay1}, when
 $X(2000)$ is treated as the $\phi(3S)$ state, the width (139 MeV) is in very good agreement with the experimental value \cite{Ablikim:2018xuz} and smaller than the theoretical result  of Ref. \cite{Barnes:2002mu}. Unfortunately, the width of the corresponding $s\bar{s}$ hybrid is in the range of $100-150$ MeV in flux tube model \cite{Page:1998gz,Close:1994hc}, which makes it difficult to determine the internal structure of this $X$ state. Table \ref{decay1} gives a comparison of the two-body decay information between $\phi(3S)$  and  $s\bar{s}$ hybrid  \cite{Page:1998gz}. Under the $\phi(3S)$ assignment, $X(2000)$ $\to KK^*(1410)\to KK\pi\pi$ will be the main decay mode with the branch ratio $\frac{\Gamma_{KK^*(1410)}}{\Gamma_{Total}}\approx 0.34$, which is smaller than that of the $s\bar{s}$ hybrid assignment.
   $KK^*$, $KK_1^\prime$ and $KK_1$ are predicted to be its important decay channels, which have the ratios of 0.2, 0.16 and 0.14, respectively.
    When treated as a $s\bar{s}$ hybrid  \cite{Page:1998gz}, $X(2000)$ dominantly decays to $KK_1^\prime$, with the branch ratio $\frac{\Gamma_{KK_1^\prime}}{\Gamma_{Total}}\approx 0.5$.
$KK^*$, $KK_1^\prime$ and $KK_1$ can decay to $KK\pi\pi$, which indicates that $KK\pi\pi$ will be the dominant final states of $X(2000)$ as the candidate of second  excitation of $\phi$. We suggest that experimentalists focus on this final channel.
 $KK$, $KK_2^*(1430)$, $K^*K^*$, and $\eta\phi$ are the sizable decay {modes} as well. 
To summarize, the branch ratios of $KK^*(1410)$, $KK_1^\prime$ and $KK$ differ greatly when $X(2000)$ is treated as $\phi(3S)$  and $s\bar{s}$ hybrid.
 These predictions of the branch ratios can help reveal the internal structure of this $X$ state.
\par
   The total width of $\phi(4S)$ is approximately 230 MeV. According to Table \ref{decay1}, the main decay modes of $\phi(4S)$ are $KK^*$, $K^*K_1^\prime$, $KK_2^*(1430)$, $KK^*(1410)$ and $KK_1$ which have the branch ratios of 0.15, 0.13, 0.12, 0.12 and  0.10, respectively. $K^*K_0^*(1430)$, $K^*K_1$  and $K^*K^*$ are its important decay channels. Considering the the final decay channels of $KK^*$, $K^*K_1^\prime$, $KK\pi$ and $KK\pi\pi$ and  will be the most important final channels in searching for the $\phi(4S)$ state experimentally.
   $KK$,  $K^*K^*(1410)$, and $KK_3^*(1780)$ also have sizable contributions to the total width of $\phi(4S)$. These predictions can help us search for and establish this $\phi(4S)$ state.
\subsubsection{D-wave $\phi$ mesons}
The  decay information of the  D-wave $\phi$ mesons is listed in Table \ref{decay2}.
\par
As shown in the second column  of Table \ref{decay2}, the strong decay of $\phi(1D)$ is predicted, which is still unobserved. $\phi(1D)$ has the total width of 547 MeV. $KK_1$ is its dominant decay channel, which {\color{black}{is consistent with}}  Ref. \cite{Wang:2012wa}. $KK^*$ and $KK$ are the important final states.   $\eta\phi$ and  $K^*K^*$  have the same ratio of 3\%.
 \begin{table}[htb]
\centering%
\caption{The partial decay widths of the $\phi$(D-wave) mesons, the unit of  widths is MeV.   \label{decay2}}
\[\begin{array}{cc cc cc}
\hline
\hline
 & \phi(\text{1D})
  & {{\text{Y(2175)}~as~\phi \text{(2D)}}}& {{\phi \text{(3D)}}}\\%
 \midrule[1pt]
Channel &Value   &Value  &Value    \\
 Total&547&{205}& {245}\\
KK_1    &423&{90.0}    &  \color{black}{67.5}\\
K^*K^* & 11.5&{ 33.4}               &\color{black}{40.5} \\
KK &40.8&{ 25.4}   & \color{black}{17.4} \\
KK^*  &57.8&{18.7}       &\color{black}{12.6} \\

 \eta\phi &13.6 & {0.879}  &\color{black}{0.3}\\
 \eta^\prime\phi&-& {0.0887}  &\color{black}{0.087 }\\
  KK^*(1410)& -&{19.6} & \color{black}{ 5.76 } \\
 KK_2^*(1430) &-& {14.5}    &\color{black}{12.1} \\

KK_1^\prime  & -&{2.56}  &\color{black}{0.59} \\

      K^*K^*(1410) &- &- &\color{black}{ 45.6} \\
       K^*K_1  &-&- & \color{black}{ 35.3 } \\
       KK_3^*(1780)&- & -&\color{black}{ 4.40} \\
       f_1(1426)\phi   &-&-&\color{black}{ 2.16 } \\
       K^*K_1^\prime  &-&- &\color{black}{ 0.454 } \\

 \hline
 \hline
\end{array}\]
\end{table}
If  $X(2000)$ is treated as the  $\phi(1D)$ state, its total width will be larger than 550 MeV, which {\color{black}{does not}}  agree with the experimental value \cite{Aubert:2006bu}. Thus, it can be basically ruled out that $X(2000)$ is the candidate of $\phi(1D)$.
 \par When treated as the $\phi(2D)$ state, $Y(2175)$ has a total width of 205 MeV, which is consistent with that in Ref.\cite{Shen:2009zze}. Under this assignment, $Y(2175)$$\to KK_1$ will be the dominant decay mode. In the calculation, $K^*K^*$ and $KK$ are the important decay channels. However, $K^*K^*$ and $KK$ modes are not observed in recent experiments\cite{Ablikim:2018xuz,Ablikim:2018iyx}. If $Y(2175)$ is the $\phi(2D)$ state, this puzzle should be explained in theory or experiment.

\par
   The decay information of the $\phi(3D)$ state is also predicted in this work.
   The total width of $\phi(3D)$  is approximately 245 MeV, with a mass of 2.6 GeV. The channels $KK_1$, $K^*K^*(1410)$ and $K^*K^*$ have the branch ratios of 0.27, 0.2 and 0.18, respectively, which are the main decay modes.
  $K^*K_1$ and $KK$ are its important decay channels. Their branch ratios are approximately 0.12 and 0.08, respectively. This work suggests that experimentalists should search for this missing state in $KK$ or $KK\pi\pi$ final states. Otherwise,
   $KK^*$ and $KK_2^*(1430)$  have sizable {\color{black}{contributions}} to the total width of $\phi(3D)$.

\section{conclusion}\label{sec4}
This paper presents an analysis of mass spectra of the excitations of $\phi$ meson, in particular the newly observed $X(2000)$ state, using the modified Godfrey-Isgur quark model, and the structure information of these excitations of the $\phi$ meson is obtained. After comparing our theoretical results of the two-body strong decays with the experimental data, we can reach the following conclusions under the conventional meson framework.
  \begin{enumerate}

\item{Mass and  strong decay behavior analysis indicates that the newly observed state
     $X(2000)$ \cite{Ablikim:2018xuz} may be the $\phi(3S)$ state, and $KK^*(1410)$ will be the dominant decay mode.}

\item{Mass analysis supports   $Y(2175)$  as a candidate of   $\phi(3S)$ or $\phi(2D)$. However, strong decay behavior analysis shows that the  $Y(2175)$ is preferably a $\phi(2D)$ state.}
\item{$\phi(4S)$ is predicted to have a mass of  2.5 GeV and a width of 230 MeV. The ground  state, $\phi(1D)$ and second radial excited state  $\phi(3D)$ have the mass {of 1.869 GeV} and  2.6 GeV and the widths of 547 MeV and 245 MeV,
     respectively}.
\end{enumerate}
\par
According to the comparison of the two-body strong decays under the $\phi(3S)$ assignment with that of the $s\bar{s}$ hybrid,  it is apparent that the study of the branch ratios of $KK^*(1410)$, $KK_1^\prime$ and $KK$ in experiment will be very valuable for identifying the nature of $X(2000)$.
\par
This study is crucial not only to establish the $\phi$ meson family and future search for the missing excitations but also to help us reveal the structure information of the newly observed $X(2000)$ state.
Thus, more experimental measurements of the resonance
parameters should be conducted by the BESIII and other
experiments, which can help us to identify the nature of $X(2000)$ and establish the $\phi$ meson family in the future.
\section{ACKNOWLEDGMENTS}
C.-Q. P. thanks Xiang Liu, Wen-biao Yan for helpful
communications and discussions. This work is supported in part by {the Nature Science Foundation Projects of Qinghai Office of Science and Technology, No. 2017-ZJ-748, the Chunhui Plan of China's Ministry of Education, No. Z2017054.}

\bibliographystyle{apsrev4-1}
\bibliography{hepref}
\end{document}